\documentstyle[aps,prl,epsf,preprint,eqsecnum]{revtex}
\newcommand{\tD}{\tilde{D}}
\newcommand{\tR}{\tilde{R}}

\newcommand{\tq}{\tilde{q}}

\newcommand{\csch}{\mbox{csch}}

\begin{document}

%%%%%%%%%%%%%%%%%%%%%%%%%%%%%%%%%%%%%%%%%%%%%%%%%%%%%%%%%%%%%%%%%%%%%%%%%%%%%
%\baselineskip 3pc
%\baselineskip 2pc
%%%%%%%%%%%%%%%%%%%%%%%%%%%%%%%%%%%%%%%%%%%%%%%%%%%%%%%%%%%%%%%%%%%%%%%%%%%%%

\preprint{gr-qc/0101065}

\draft

\begin{titlepage}
  \title{
  Initial condition of a gravitating thick loop cosmic string
  and linear perturbations 
  }
  \author{
  Kouji NAKAMURA\footnote{E-mail address : kouchan@phys-h.keio.ac.jp}
  }
  \address{
  Theory Division, National Astronomical Observatory, Mitaka, Tokyo, 181-8588,
  Japan \\
  and\\
  Department of Physics, Keio University,
  Hiyoshi Yokohama, 223-8521, Japan.
  }
  \date{\today}
\maketitle

\begin{abstract}
  The initial data of the gravitational field produced by a loop
  thick string is considered.
  We show that a thick loop is not a geodesic on the initial
  hypersurface, while a loop conical singularity is.
  This suggests that there is the ``{\it critical thickness}''
  of a string, at which the linear perturbation theory with a
  flat space background fails to describe the gravity of a loop
  cosmic string.
  Using the above initial data, we also show that the linear
  perturbation around flat space is plausible if the
  string thickness is larger than $\sim5\times10^{-3}a$, where
  $a$ is the curvature radius of the loop. 
\end{abstract}
\end{titlepage}

%\begin{multicols}{2}

%%%%%%%%%%%%%%%%%%%%%%%%%%%%%%%%%%%%%%%%%%%%%%%%%%%%%%%%%%%%%%%%%%%%%%%
\section{Introduction}
\label{sec:Introduction}
%%%%%%%%%%%%%%%%%%%%%%%%%%%%%%%%%%%%%%%%%%%%%%%%%%%%%%%%%%%%%%%%%%%%%%%

For the past few decades, gravity of cosmic strings and
gravitational waves from them have been investigated by many 
researchers\cite{Vilenkin_Shellard}--\cite{Boisseau}. 
There are two physical contexts to investigate gravity of cosmic
strings.
One is the cosmological context and another is in the technical
problem in general relativity: the treatment of concentrated
line sources.

%*************************************************************

In the cosmological context, cosmic strings are topological
defects associated with the symmetry breaking in unified
theories\cite{Vilenkin_Shellard}. 
In the simplest case, the dynamics of cosmic strings is governed
by the Nambu-Goto action. 
If the self-gravity of strings is ignored (test string case),
the Nambu-Goto action admits oscillatory solutions. 
It is considered that these oscillations lead the emission of
gravitational waves and strings gradually lose their kinetic
energy by this emission\cite{Vachaspnati}. 
Further, relic gravitational waves emitted by cosmic strings
might be detectable by the gravitational wave detectors to be
constructed in the near future\cite{Caldwell-Battye-Shellard}. 
Though this is a quite fascinating prediction, it is not clear
whether or not it can be taken seriously.
Recently, by taking into account the self-gravity of the
Nambu-Goto string, it is shown that the oscillatory behavior of
a self-gravitating string coupled to gravitational wave is quite
different from that of a test string\cite{kouchan}. 
The dynamical degree of freedom concerning the perturbative
oscillations of an infinite string is completely determined by
that of gravitational waves and a self-gravitating infinite
string does not oscillate spontaneously unlike an infinite test
string.

%*************************************************************

To give a completely precise dynamics of gravitating cosmic
strings, we must solve the Einstein equations and the evolution
equations of the Higgs and gauge fields from generic initial
data, which is impossible at present.
So it will be instructive to replace a cosmic string by a
concentrated line source and clarify its gravity at
first\cite{line-sources}.
However, it is known that the mathematical description of a
self-gravitating thin string is delicate because the support of
a string is a surface of co-dimension two.
In general relativity, there is no simple prescription of an
arbitrary concentrated line source where the metric becomes
singular\cite{Geroch}.

%*************************************************************

In spite of these difficulties, a conical singularity, which
appears in some exact solutions to the Einstein equations, is
often regarded as a self-gravitating cosmic string. 
(For example, see Ref.\cite{Kinnersley-Walker}.)
This is one of the idealizations of the concentrated matter
sources.
Since a spacetime containing a straight gauge string has
the asymptotically conical structure\cite{conical}, this
idealization seems plausible when the string thickness is much
smaller than the other physical scale of interests.

%*************************************************************

As an extension of this idealization, one might expect that an
oscillating cosmic string can be also idealized by an
oscillating conical singularity. 
However, as pointed out by Unruh et
al.\cite{Unruh-Hayward-Israel-McManus}, the worldsheet of a
conical singularity is totally geodesic.
This behavior is also seen in the time symmetric initial data of
gravity of a loop conical singularity\cite{Frolov-Israel-Unrhu}
and similar results are also obtained in
Refs.\cite{kouchan,Boisseau}.
The dynamics of conical singularities is completely determined
without their equations of motion. 
This geodesic property of a conical singularity shows that the
idealization of cosmic strings by conical singularities might
gives the different dynamics from that of cosmic strings in the
early universe which have their finite thickness.
Further, this implies that the dynamics of cosmic strings is not 
idealized by that of conical singularities, though it seems
natural to expect that loops of cosmic string can be regarded
as conical singularities when the loop radius is much larger
than the string thickness.

%*************************************************************

In this paper, we reconsider the above geodesic property pointed
out in Ref.\cite{Unruh-Hayward-Israel-McManus} using a time
symmetric initial data containing a thick loop string.
We show that the zero thickness of strings is essential to
this geodesic property. 
Because of the singular metric, gravity near the loop conical
singularity is so strong that the loop itself is a geodesic on 
the initial surface. 
On the other hand, the core of the thick loop string is not a
geodesic on the initial surface and the curvature of a test loop
on a flat space is naturally reproduced in the limit
$G\rightarrow 0$ where $G$ is Newton's gravitational constant.
Then, the geodesic property in
Ref.\cite{Unruh-Hayward-Israel-McManus} should not be taken
seriously when we consider the dynamics of cosmic strings, since
cosmic strings in the early universe have their own finite
thickness.
This also implies that we must take into account the effect of
the string thickness when we discuss the dynamics of cosmic strings.

%*************************************************************

This behavior of the gravity of a thick string loop also
suggests that there is a criterion in the string thickness at
which the linear perturbation theory around a flat background
spacetime fails to describe the gravity of cosmic strings.
We call this thickness ``{\it critical thickness}'' and evaluate 
this criterion in this paper. 
The geodesic property in Ref.\cite{Unruh-Hayward-Israel-McManus}
does not depend on the loop radius and its deficit angle.
When the spacetime contains a sufficiently small loop conical
singularity, this loop is a geodesic on this spacetime and the
other geodesics on this spacetime are quite different from those
on the flat spacetime, on which geodesics are straight lines. 
On the other hand, gravity of a thick loop string is not so
strong as mentioned above. 
This suggests that the linear gravity is plausible if the string
is sufficiently thick. 
Then, for a fixed loop radius and a fixed deficit angle, there
exists a critical thickness and the linear gravity on a flat
background spacetime is valid if the string thickness is much
greater than this critical thickness.

%*************************************************************

To discuss the initial data of the gravity of a thick string
loop and evaluate the above critical thickness, we first
construct an example of the initial data in which the matter
concentrates on the torus shell. 
Since the gravity of a cosmic string has asymptotically conical
structure, we use the initial data derived by Frolov et
al.\cite{Frolov-Israel-Unrhu} as a solution outside the torus.
Using this example, we explicitly show that the core of the loop
string is not a geodesic if the string has its finite thickness,
while the loop approaches to a geodesic in the zero-thickness
limit.
Though the matter distribution in this example is too artificial
as a realistic cosmic string loop, our conclusion is also true
even if the matter distribution is smooth. 
Our example also shows that the initial data derived by Frolov
et al.\cite{Frolov-Israel-Unrhu} may be regarded as the
gravitational field produced by a smooth matter distribution.
So, using this initial data as the geometry outside the matter
field, we derive the above critical thickness of a string by
comparing with a linear solution on the flat space background.

%*************************************************************

This paper is organized as follows: In
Sec.\ref{sec:general-formalizm}, we briefly review the ADM
initial constraints and the treatment of the $\delta$-function
distribution of the matter field on the initial surface to
construct an example of the gravitational field produced by a
loop matter distribution. 
In Sec.\ref{sec:loop-thick-string-initial}, we construct an
initial data of a thick loop string and discuss whether the
string is a geodesic on the initial surface or not. 
We discuss the above critical thickness in
Sec.\ref{sec:liniear}.  
The final section (Sec.\ref{sec:disucssion}) is devoted to
summary and discussions.

%*************************************************************

Throughout this paper we work in relativistic units, so that the
speed of light is equal to unity.

%*************************************************************

%%%%%%%%%%%%%%%%%%%%%%%%%%%%%%%%%%%%%%%%%%%%%%%%%%%%%%%%%%%%%%%%%%%%%%%
\section{Initial value constraints and Junction conditions}
\label{sec:general-formalizm}
%%%%%%%%%%%%%%%%%%%%%%%%%%%%%%%%%%%%%%%%%%%%%%%%%%%%%%%%%%%%%%%%%%%%%%%

In 4-dimensional spacetime $({\cal M},g_{ab})$, we consider a
zero extrinsic curvature 3-dimensional spacelike submanifold
$(\Sigma,q_{ab})$. 
We denote the unit normal vector of $\Sigma$ in ${\cal M}$ by
$u^{a}$. 
The induced metric $q_{ab}$ on $\Sigma$ from the metric $g_{ab}$
on ${\cal M}$ is given by $q_{ab} := g_{ab} + u_{a}u_{b}$. 
The zero extrinsic curvature means the spacetime is momentarily
static at $(\Sigma,q_{ab})$. 
The ADM initial-value constraints for a momentarily static
hypersurface $(\Sigma,q_{ab})$ are reduced 
to\cite{Wald}  
\begin{equation}
  \label{Hamiltonian_constraint-1}
  {}^{(3)}\!R = 16\pi G T_{\perp\perp},
\end{equation}
where ${}^{(3)}\!R$ is the Ricci scalar curvature of $\Sigma$
and $T_{\perp\perp}:=T_{ab}u^{a}u^{b}$ is the energy density of
the matter field on the hypersurface $\Sigma$. 
In terms of a conformally transformed metric $\tq_{ab} :=
\phi^{-4} q_{ab}$, the constraint
(\ref{Hamiltonian_constraint-1}) is given by   
\begin{equation}
  \label{hamiltonian-constraint-2}
  \tD^{a} \tD_{a}\phi 
  - \frac{1}{8} \phi {}^{(3)}\!\tR 
  = - 2\pi G \phi^{5} T_{\perp\perp}, 
\end{equation}
where $\tD_{a}$ and ${}^{(3)}\!\tR$ are the covariant derivative
and the scalar curvature associated with the metric $\tq_{ab}$,
respectively. 
In this paper, we concentrate only on the geometry of $\Sigma$
using the constraint (\ref{hamiltonian-constraint-2}) 
(or (\ref{Hamiltonian_constraint-1})).

%*************************************************************

For simplicity, we consider $\delta$-function distribution of
matter energy density at the 2-dimensional hypersurface
${\cal S}$ on $\Sigma$ to construct an example of solutions to
Eq.(\ref{hamiltonian-constraint-2}). 
To treat $\delta$-function distribution of the matter energy, we
apply Israel's condition\cite{Israel} on the initial surface,
which is derived from the constraint
(\ref{hamiltonian-constraint-2}). 
To derive this condition, we first divide $(\Sigma,q_{ab})$ into
two manifolds $(\Sigma_{\pm},q_{ab\pm})$ with the boundaries
$\partial\Sigma_{\pm}$ and then identify ${\cal
S}=\partial\Sigma_{+}=\partial\Sigma_{-}$. 
This identification implies $n_{a+}=n_{b-}$, where $n_{a\pm}$
are the unit normal vector of $\partial\Sigma_{\pm}$ on
$\Sigma_{\pm}$, respectively. 
The induced metrics $h_{ab\pm}$ on $\partial\Sigma_{\pm}$ from
$q_{ab\pm}$ are given by
$h_{ab\pm}:=q_{ab\pm}-n_{a\pm}n_{b\pm}$, respectively. 
Since $\partial\Sigma_{\pm}$ are diffeomorphic to each other,
$h_{ab\pm}$ satisfy the condition
\begin{equation}
  \label{constraint-junction-intrinsic}
  h_{ab+} = h_{ab-}.
\end{equation}

%*************************************************************

By virtue of the constraint (\ref{Hamiltonian_constraint-1}),
the $\delta$-function distribution of the matter energy density
also gives the gap of the extrinsic curvatures $\kappa_{ab\pm}$
of $\partial\Sigma_{\pm}$ facing to $\Sigma_{\pm}$.
$\kappa_{ab\pm}$ are defined by $\kappa_{ab\pm} := -
(h_{a}^{\;\;c}h_{b}^{\;\;d}D_{c}n_{d})_{\pm}$, where $D_{a}$ is
the covariant derivative associated with the metric $q_{ab}$. 
Introducing the Gaussian normal coordinate $\chi$ in the
neighborhood of ${\cal S}$ by $n_{a\pm} = (d\chi)_{a}$, the
scalar curvature ${}^{(3)}\!R$ on $\Sigma$ is decomposed as follows:   
\begin{equation}
  {}^{(3)}\!R = {}^{(2)}\!R
  - (\kappa^{\;\;a}_{a})^{2} + \kappa_{a}^{\;\;b} \kappa_{b}^{\;\;a}
  + 2 \frac{\partial}{\partial\chi} \kappa_{a}^{\;\;a}.
\end{equation}
When $T_{\perp\perp}$ is given by
\begin{equation}
  \label{Tpp-def}
  T_{\perp\perp} = \mu \delta(\chi),
\end{equation}
the integration of (\ref{Hamiltonian_constraint-1}) over the
infinitesimal interval of $\chi$ yields  
\begin{equation}
  \label{constraint-junction-extrinsic}
  \kappa_{a+}^{\;\;a} - \kappa_{a-}^{\;\;a} = 8\pi G \mu.
\end{equation}
Hence all conditions for the identification ${\cal
S}=\partial\Sigma_{\pm}$ on the momentarily static initial data
are (\ref{constraint-junction-intrinsic}) and 
(\ref{constraint-junction-extrinsic}).

%*************************************************************

%%%%%%%%%%%%%%%%%%%%%%%%%%%%%%%%%%%%%%%%%%%%%%%%%%%%%%%%%%%%%%%%%%%%%%%
\section{Initial condition of a thick loop string}
\label{sec:loop-thick-string-initial}
%%%%%%%%%%%%%%%%%%%%%%%%%%%%%%%%%%%%%%%%%%%%%%%%%%%%%%%%%%%%%%%%%%%%%%%

%*************************************************************

Here, we consider a momentarily static initial data of gravity
produced by a thick loop string. 
In this paper, we assume that the vacuum region outside the loop
is given by the initial data for a loop conical singularity
derived by Frolov et al.\cite{Frolov-Israel-Unrhu} as mentioned
in the Introduction (Sec.\ref{sec:Introduction}).
In this section, we first review the conformally flat version of
the initial data for a loop conical singularity in
Ref.\cite{Frolov-Israel-Unrhu}.  
Second, replacing the conical singularity by a matter
distribution, we construct an example of a gravitating thick
loop string. 
Then we discuss whether the loop is a geodesic on the initial
surface or not. 
Though we concentrate only on a conformally flat initial data,
the similar behavior is also seen in the conformally non-flat
version. (See Appendix \ref{sec:c-non-flat}.)

%*************************************************************

%%%%%%%%%%%%%%%%%%%%%%%%%%%%%%%%%%%%%%%%%%%%%%%%%%%%%%%%%%%%%%%%%%%%%%%
\subsection{Initial condition of a loop conical singularity}
\label{sec:conical-singularity-initial}
%%%%%%%%%%%%%%%%%%%%%%%%%%%%%%%%%%%%%%%%%%%%%%%%%%%%%%%%%%%%%%%%%%%%%%%

\begin{figure}[htbp]
  \begin{center} 
    \leavevmode
    \epsfxsize=0.55\textwidth
    \epsfbox[0 0 2095 1559]{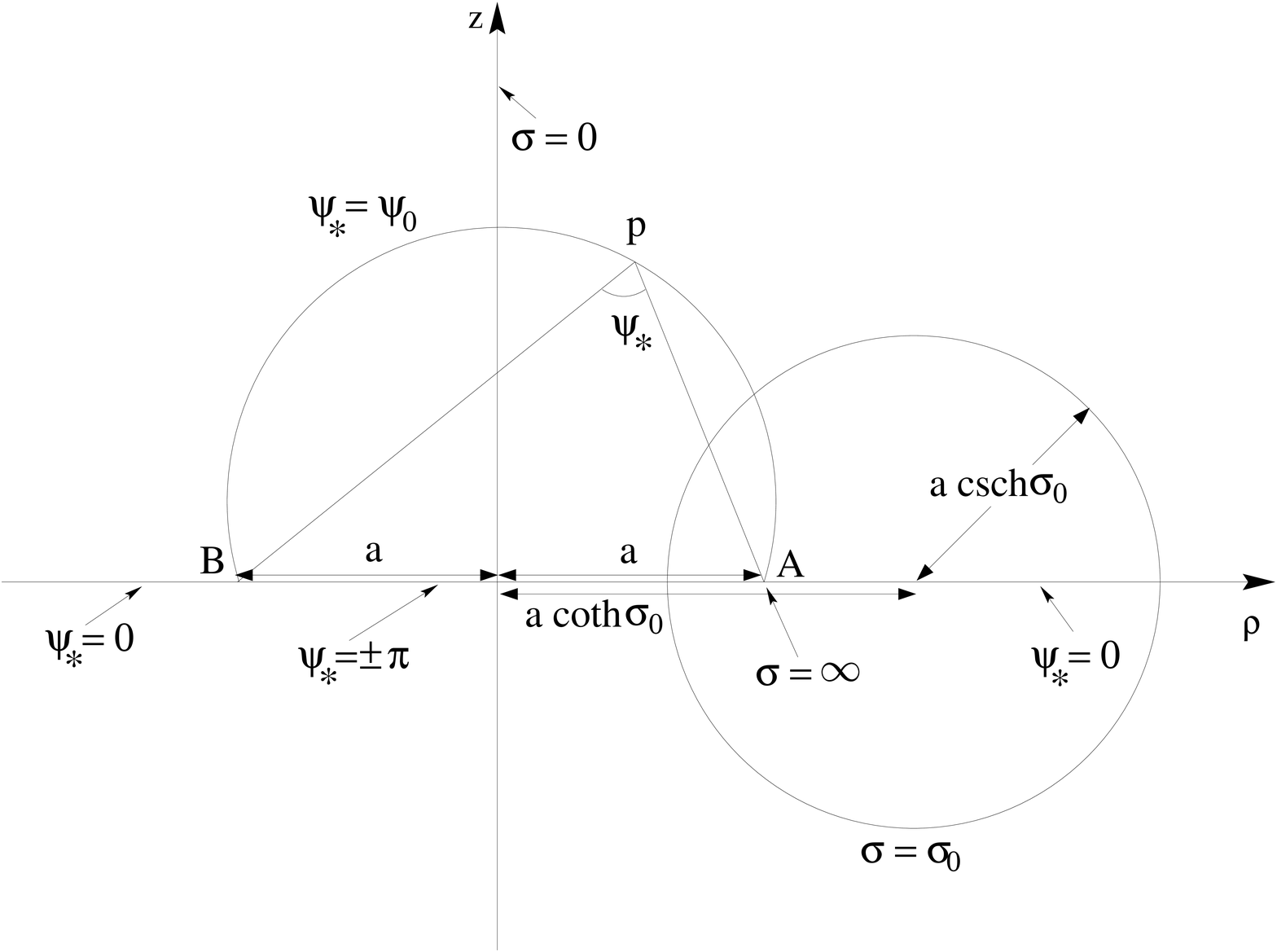}
    \caption{
    Toroidal coordinates $\sigma$, $\psi_{*}$ is defined on the
    extension of a $\varphi$=constant half-plane in Euclidean
    space. 
    $A$ and $B$ are the points at which the equatorial circle
    $\rho=a$ intersects this plane. 
    For any point $P(\sigma,\psi_{*},\varphi)$,
    $\sigma=\ln(AP/PB)$, and $\psi_{*}=\pm$(angle $APB$), with
    the sign equal to the sign of $z$. 
    }  
    \label{fig:fig1}
  \end{center}
\end{figure}

To discuss the initial condition for a loop conical singularity,
we introduce the toroidal coordinate system. 
First, we consider the flat space metric 
\begin{equation}
  \label{Eucl-cylindrical}
  ds^{2}_{E} = d\rho^{2} + dz^{2} + \rho^{2}d\varphi^{2}.
\end{equation}
Defining the functions $\sigma$ and $\psi_{*}$ by 
\begin{equation}
  \rho = a \frac{\sinh\sigma}{N^{2}}, \quad   
  z = a \frac{\sin\psi_{*}}{N^{2}}, \quad 
  N(\sigma,\psi_{*}) := \sqrt{\cosh\sigma-\cos\psi_{*}},
\end{equation}
where $0\leq\sigma\leq\infty$ and $-\pi \leq \psi_{*} \leq\pi$,
the line element (\ref{Eucl-cylindrical}) becomes
\begin{equation}
  \label{flat-toroidal}
    ds^{2}_{E} = 
    \frac{1}{N^{4}} ds^{2}_{*} := 
    \frac{a^{2}}{N^{4}} \left(d\sigma^{2} + d\psi_{*}^{2} +
      \sinh^{2}\sigma d\varphi^{2}\right).
\end{equation}
The geometrical meaning of the coordinates $\sigma$ and
$\psi_{*}$ is depicted in Fig.\ref{fig:fig1}. 
A $\sigma$=constant surface is a torus with circular cross
sections of radius $a\csch\sigma$; the central axis of the tube
forms a circle of radius $\rho=a\coth\sigma$ in the plane
$z=0$. 
The ring $\rho=a$, $z=0$ is the limit torus $\sigma=\infty$.

%*************************************************************

\begin{figure}[htbp]
  \begin{center}
    \leavevmode
    \epsfxsize=0.55\textwidth
    \epsfbox[0 -100 1671 1274]{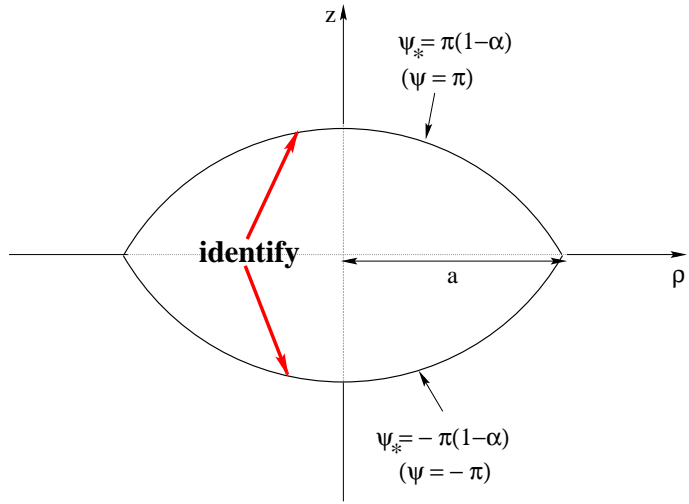}
    \caption{
    The conformally flat geometry of the circular string loop,
    located at $\rho=a$ in this Euclidean map, is constructed by
    conformally squeezing the two spherical caps
    $\psi_{*}=\pm\pi(1-\alpha)$ to become in effect, equatorial
    disks spanning the loop, and topologically identifying them.
    } 
    \label{fig:fig2}
  \end{center}
\end{figure}

To construct the initial data of a loop conical singularity at
the limit torus $\sigma=\infty$, we remove a pair of spherical
caps spanning the loop as shown in
Fig.\ref{fig:fig2}\cite{Frolov-Israel-Unrhu}.
First, we fix an arbitrary angular deficit $2\pi\alpha<2\pi$. 
We restrict $\psi_{*}$ to the range
$-\pi(1-\alpha)<\psi_{*}<\pi(1-\alpha)$, identify these end
points of this interval. 
Applying a conformal factor $(\phi N)^{4}$ to
Eq.(\ref{flat-toroidal}), the physical line element associated
with the metric $q_{ab}$ is given by
\begin{equation}
  \label{physical-three-metric}
  ds^{2} = (\phi N)^{4}ds^{2}_{E} = \phi^{4}ds^{2}_{*}.
\end{equation}
Second, we rescale the coordinate $\psi_{*}$ by  
\begin{equation}
  \psi := \eta\psi_{*}, \quad -\pi\leq\psi\leq\pi, 
  \quad \eta:=\frac{1}{1-\alpha}.
\end{equation}
Then the conformally transformed metric $\tq_{ab}$ is given by 
\begin{equation}
  \label{conformally-transformed-metric-outside}
  \tq_{ab} = a^{2}\left((d\sigma)_{a}(d\sigma)_{b} 
    + \frac{1}{\eta^{2}}(d\psi)_{a}(d\psi)_{b} 
    + \sinh^{2}\sigma (d\varphi)_{a}(d\varphi)_{b} \right).
\end{equation}

%*************************************************************

Though the exposed spherical faces have the same intrinsic
geometry, their extrinsic curvatures have opposite sign in
$(\Sigma,N^{-4}\tq_{ab})$.
This discontinuity disappears in $(\Sigma,q_{ab})$ if the
conformal factor $\phi(\sigma,\psi)$ satisfies the following
boundary conditions:
\begin{enumerate}
  \renewcommand{\labelenumi}{(\roman{enumi})}
\item $\phi(\sigma,\psi)$ is even in the inversion
  $\psi\rightarrow -\psi$ and smooth at $\psi=\pm\pi$;
\item $\phi N\rightarrow 1$ at $\sigma\rightarrow\infty$;
\item $\phi N\rightarrow$ constant at $\sigma\rightarrow 0$.
\end{enumerate}
By virtue of the condition (i), the extrinsic curvature of the
surface $\psi=\pm\pi$ in $(\Sigma,q_{ab})$ vanishes and no
discontinuity is there. 
The condition (ii) guarantees that $(\Sigma,q_{ab})$ has the
conical structure with the deficit angle $\alpha$ at the limit
tours $\sigma=\infty$. 
$(\Sigma,q_{ab})$ is asymptotically flat by virtue of the final
condition (iii).

%*************************************************************

Assuming $\varphi^{a}:=(\partial/\partial\varphi)^{a}$ is a
Killing vector on $(\Sigma,q_{ab})$,
Eq.(\ref{hamiltonian-constraint-2}) for the vacuum region
$\Sigma_{-}$ in the coordinate system $(\sigma,\psi,\varphi)$ is
given by 
\begin{equation}
  \label{hamiltonian-constraint-vacuum}
  \frac{1}{\sinh\sigma}\partial_{\sigma}
  \left(\sinh\sigma\partial_{\sigma}\phi\right) +
  \eta^{2}\partial_{\psi}^{2}\phi + \frac{1}{4}\phi = 0. 
\end{equation}
The general solution to this equation with the boundary
conditions (i) and (ii) has the form  
\begin{equation}
  \label{vacuum-solution-general}
  \phi = \frac{\sqrt{2}}{\pi} \sum_{n=0}^{\infty}
  a_{n}\epsilon_{n}Q_{\eta n-\frac{1}{2}} (\cosh\sigma)\cos n\psi,
\end{equation}
where $Q_{\nu}(*)$ is the second class Legendre function and
$\epsilon_{n}:=2-\delta_{n}^{0}$ is the Neumann factor. 
By comparing this with the expansion  
\begin{equation}
  \label{N-expand}
  N^{-1}(\sigma,\psi_{*}) = \frac{\sqrt{2}}{\pi} \sum^{\infty}_{n=0}
  \epsilon_{n} Q_{n-\frac{1}{2}} (\cosh\sigma)\cos n\psi_{*},
\end{equation}
one can easily see that 
\begin{equation}
  \label{vacuum-solution-1}
  \phi = \frac{\sqrt{2}}{\pi}
  \sum^{\infty}_{n=0} \epsilon_{n}
  Q_{\eta n-\frac{1}{2}}(\cosh\sigma)\cos n\psi, 
\end{equation}
is one of the simplest choice which continuously reduces to
Eq.(\ref{N-expand}) in the limit $\alpha\rightarrow 0$ and
satisfies the boundary conditions (i) and (ii).

%*************************************************************

From the asymptotic behavior of Eq.(\ref{vacuum-solution-1}) in
the limit
\begin{equation}
  r:=\sqrt{\rho^{2} + z^{2}}=\frac{a}{N}\sqrt{\cosh\sigma +
  \cos\psi} \rightarrow \infty,
\end{equation}
we can easily confirm that $(\Sigma,q_{ab})$ is asymptotically
flat, i.e., Eq.(\ref{vacuum-solution-1}) satisfies the boundary
condition (iii). 
Since 
\begin{equation}
  \label{eta-phi-N-asympt}
  \eta\phi N \sim 1 + \frac{2a}{\pi} \frac{I(\eta)}{\sqrt{\rho^{2} + z^{2}}}  
  = 1 + \frac{2a}{\pi} \frac{I(\eta)}{r}, \quad (r\rightarrow\infty), 
\end{equation}
where
\begin{equation}
  I(\eta) := \int^{\infty}_{0}\csch x (\eta\coth\eta x - \coth x)dx,
\end{equation}
the physical line element (\ref{physical-three-metric}) is given by 
\begin{equation}
  \label{asymptotic-metric}
  ds^{2} = (\eta \phi N)^{4} \frac{ds^{2}_{E}}{\eta^{4}} 
  \sim 
  \left(1 + \frac{8a}{\pi} \frac{I(\eta)}{r}\right)
  \left\{ 
    \left(\frac{dr}{\eta^{2}}\right)^{2} 
    + \left(\frac{r}{\eta^{2}}\right)^{2} d\Omega_{2}^{2} \right\}.
\end{equation}
This shows that $(\Sigma,q_{ab})$ has asymptotically flat region
with the ADM mass 
\begin{equation}
  M_{ADM} = \frac{4a}{\pi\eta^{2}} I(\eta).
\end{equation}

%*************************************************************

%%%%%%%%%%%%%%%%%%%%%%%%%%%%%%%%%%%%%%%%%%%%%%%%%%%%%%%%%%%%%%%%%%%%%%%
\subsection{Thick string initial data}
\label{sec:thick-initial}
%%%%%%%%%%%%%%%%%%%%%%%%%%%%%%%%%%%%%%%%%%%%%%%%%%%%%%%%%%%%%%%%%%%%%%%

%*************************************************************

Here, we replace the above conical singularity by a matter
distribution. 
For simplicity, we consider an example in which the matter
concentrates on the torus ${\cal S}$ ($\sigma$=constant) as
Eq.(\ref{Tpp-def}), where $\mu$ may depend on $\psi$.
We denote the outside the matter torus by $(\Sigma_{-},q_{ab-})$
and the inside by $(\Sigma_{+},q_{ab+})$.

%*************************************************************

$(\Sigma_{-},q_{ab-})$ includes asymptotically flat region and
$q_{ab-}$ is given by Eq.(\ref{physical-three-metric}) with the
conformal factor (\ref{vacuum-solution-1}). 
Since we assume that $(\Sigma_{+},q_{ab+})$ is vacuum without
conical singularities, the metric $q_{ab+}$ is given by  
\begin{equation}
  \label{inside-physical-metric}
  q_{ab+} = a^{2} \phi^{4} \left((d\sigma)_{a}(d\sigma)_{b} 
    + (d\psi)_{a}(d\psi)_{b} 
    + \sinh^{2}\sigma (d\varphi)_{a}(d\varphi)_{b}\right),
\end{equation}
where $\phi$ is a solution to
Eq.(\ref{hamiltonian-constraint-vacuum}) with $\eta=1$.  
Imposing the boundary conditions (i), (ii) and $\eta=1$, the
solution to Eq.(\ref{hamiltonian-constraint-vacuum}) is given by
\begin{equation}
  \label{vacuum-solution-inside}
  \phi = \frac{\sqrt{2}}{\pi} \sum_{n=0}^{\infty}
  a_{n}\epsilon_{n}Q_{n-\frac{1}{2}} (\cosh\sigma)\cos n\psi.
\end{equation}
The coefficients $a_{n}$ in Eq.(\ref{vacuum-solution-inside})
are determined by the boundary condition
(\ref{constraint-junction-intrinsic}) at ${\cal S}$.

%*************************************************************

Here, we consider the identification ${\cal
S}=\partial\Sigma_{\pm}$ at the torus shell. 
We assume that $\varphi^{a}$ is a Killing vector on the whole
initial surface $\Sigma$ and identify the orbits of
$\varphi^{a}$ at ${\cal S}=\partial\Sigma_{\pm}$. 
Since $\psi$ is a periodic coordinate with the period $2\pi$ 
on each $\Sigma_{\pm}$, we also identify the orbits of
$\psi^{a}=(\partial/\partial\psi)^{a}$ at ${\cal
S}=\partial\Sigma_{\pm}$. 
By these identifications, the functions $\varphi$ and $\psi$ are
extended to smooth functions on $\Sigma$. 
Finally, we choose the coordinate function $\sigma$ so that
$\sigma=\sigma_{1\pm}$ represent $\partial\Sigma_{\pm}$ in
$\Sigma_{\pm}$, respectively.

%*************************************************************

The intrinsic metrics $h_{ab\pm}$ on $\partial\Sigma_{\pm}$ are
given by
\begin{equation}
  \begin{array}{l}
  h_{ab-} = a^{2} \phi^{4}_{-}(\psi) 
  \left( \frac{\displaystyle 1}{\displaystyle \eta^{2}}(d\psi)_{a}(d\psi)_{b}
    + \sinh^{2}\sigma_{1-} (d\varphi)_{a}(d\varphi)_{b}\right), \\
  h_{ab+} = a^{2} \phi^{4}_{+}(\psi) 
  \left( (d\psi)_{a}(d\psi)_{b} + \sinh^{2}\sigma_{1+}
  (d\varphi)_{a}(d\varphi)_{b}\right), 
  \end{array}
\end{equation}
where $\phi_{\pm} := \phi(\sigma=\sigma_{1\pm})$, respectively. 
Then, Eq. (\ref{constraint-junction-intrinsic}) yields
\begin{equation}
  \label{intrinsic-junctions}
  \eta \sinh\sigma_{1-} = \sinh\sigma_{1+}, \quad
  \phi_{+} = \phi_{-} \eta^{-\frac{1}{2}}.
\end{equation}
The first equation in Eqs.(\ref{intrinsic-junctions})
determineds $\sigma_{1+}$ as a function of $\sigma_{1-}$ and the
deficit angle $\alpha=1-1/\eta$. 
The second one in Eqs.(\ref{intrinsic-junctions})
determines the coefficients $a_{n}$ in $\phi_{+}$: 
\begin{equation}
  \label{a_n-relation}
  a_{n} = \frac{ \eta^{-\frac{1}{2}} Q_{n\eta-\frac{1}{2}}(\cosh\sigma_{1-})}
  { Q_{n-\frac{1}{2}}(\cosh\sigma_{1+})}. 
\end{equation}

%*************************************************************

Israel's junction (\ref{constraint-junction-extrinsic}) determines
the energy density $\mu$ as a function of $\psi$: 
\begin{equation}
  8\pi G\mu(\psi) = - \frac{4}{a\phi^{3}_{+}(\psi)}
  \partial_{\sigma}\phi_{+}(\psi) 
  - \frac{\coth\sigma_{1+}}{a\phi^{2}_{+}(\psi)}
  + \frac{4}{a\phi^{3}_{-}(\psi)} 
  \partial_{\sigma}\phi_{-}(\psi)
  + \frac{\coth\sigma_{1-}}{a\phi^{2}_{-}(\psi)}.
\end{equation}
When the matter torus is sufficiently thin, i.e.,
$\sigma_{1\pm}\rightarrow\infty$, $\phi_{\pm}$ behave 
\begin{equation}
  \label{phi+phi-asympt}
  \begin{array}{rcl}
    \phi_{+} &=& \sqrt{2}e^{-\frac{1}{2}\sigma_{1+}} \left(a_{0} +
      a_{1}e^{-\sigma_{1+}}\cos\psi + O(e^{-2\sigma_{1+}})\right),\\
    \phi_{-} &=& \sqrt{2}e^{-\frac{1}{2}\sigma_{1-}} \left(1 +
      \frac{\displaystyle 2\Gamma(\eta+1/2)}
      {\displaystyle \sqrt{\pi}\Gamma(\eta+1)}
      e^{-\eta\sigma_{1-}}\cos\psi + O(e^{-2\sigma_{1-}})\right). 
  \end{array}
\end{equation}
From Eqs.(\ref{phi+phi-asympt}) and Eq.(\ref{a_n-relation}), we
easily see that
\begin{eqnarray}
  a\phi^{2}_{+}(\sigma_{1+},\psi)\mu(\psi) 
  = \frac{1}{8\pi G} \left[
    \left(1 - \frac{1}{\eta}\right) + O(e^{-2\sigma_{1\pm}})
  \right].
  \label{rho-asymptotic-form-of-flat-2}
\end{eqnarray}
This shows that $a\phi^{2}_{+}\mu(\psi)$ does not depend $\psi$
up to the order of $O(e^{-2\sigma_{1\pm}})$. 
The leading term is just the relation of the proper line energy
density of a straight string and the deficit angle.

%*************************************************************

%%%%%%%%%%%%%%%%%%%%%%%%%%%%%%%%%%%%%%%%%%%%%%%%%%%%%%%%%%%%%%%%%%%%%%%
\subsection{Is a loop string is a geodesic?}
\label{sec:is-a-loop-string-is-geodesic}

If the solution (\ref{vacuum-solution-1}) is extended to the
limit torus $\sigma=\infty$, $(\Sigma,q_{ab})$ contains a loop
conical singularity at the limit torus. 
This conical singularity is a geodesic on $(\Sigma,q_{ab})$ as
pointed out by Unruh et al.\cite{Unruh-Hayward-Israel-McManus}. 
To see this from Eq.(\ref{vacuum-solution-1}), we consider the
orbit of the Killing vector $\varphi^{a}$ and pay attention to
the magnitude of the ``acceleration'' of this Killing orbit. 
In the neighborhood of the limit torus $\sigma=\infty$, the norm
$V$ of $\varphi^{a}$ behaves 
\begin{equation}
  \frac{V^{2}}{a^2} = \phi^4 \sinh^{2}\sigma 
  = 1 + \frac{8}{\sqrt{\pi}} e^{- \eta \sigma} 
  \frac{\Gamma(\eta + \frac{1}{2})}{\Gamma(\eta+1)}
  \cos\psi + O(e^{-2\sigma}).  
\end{equation}
Then the magnitude $\kappa$ of the ``acceleration'' is given by 
\begin{equation}
  \label{Frolov-Israel-Unrhu-kappa}
  \kappa = \frac{\sqrt{(D^{a}V)(D_{a}V)}}{V}
 \sim \frac{2\eta}{\sqrt{\pi}a} 
  \frac{\Gamma(\eta+\frac{1}{2})}{\Gamma(\eta+1)} e^{- (\eta - 1)\sigma} 
  \sim \frac{2\eta}{\sqrt{\pi}a} 
  \frac{\Gamma(\eta+\frac{1}{2})}{\Gamma(\eta+1)}
  \left(\frac{d}{2a}\right)^{\eta - 1},  
\end{equation}
where $d\sim 2a e^{-\sigma}$ is the proper distance from the
Killing orbits of $\varphi^{a}$ with a finite $\sigma$ to the
limit torus $\sigma=\infty$. 
Eq.(\ref{Frolov-Israel-Unrhu-kappa}) shows that $\kappa$
vanishes when $d\rightarrow 0$. 
Then the loop conical singularity is a geodesic on
$(\Sigma,q_{ab})$.
We note that this is also true in the initial data of the
conformally non-flat version derived in
Ref.\cite{Frolov-Israel-Unrhu}. 
(See Appendix \ref{sec:c-non-flat}.)

%*************************************************************

Here, we show that the core of a thick loop string (the limit
torus $\sigma=\infty$) is not a geodesic on the initial surface
$(\Sigma,q_{ab})$ containing a thick loop string in
Sec.\ref{sec:thick-initial}.
In this initial data, the ``acceleration'' $\kappa$ of the
Killing orbit of $\varphi^{a}$ behaves 
\begin{equation}
  \label{kappa-vacuum}
  \kappa = \frac{a_{1}}{a a_{0}^{3}} + O(e^{-\sigma})
\end{equation}
in the limit $\sigma\rightarrow\infty$. 
$a_{0}$ and $a_{1}$ are given by Eqs.(\ref{a_n-relation}) and
are constants determined by the outside deficit angle $\alpha$
and the locus $\sigma=\sigma_{1+}$ (or $\sigma_{1-}$) of the
matter torus. 
Thus, the inside matter torus shell, the orbit of $\varphi^{a}$
is not a geodesic on the whole initial surface $\Sigma$ if a
string has its finite thickness.

%*************************************************************

When the torus shell of the matter field is sufficiently thin
($\sigma_{1\pm}\rightarrow\infty$), $\kappa$ behaves 
\begin{equation}
  \label{kappa-inside-thin-string}
  \kappa 
  = \frac{1}{a}\frac{2}{\sqrt{\pi}}\frac{\Gamma(\eta+1/2)}
  {\Gamma(\eta+1)}
  \eta^{\eta} e^{-(\eta-1)\sigma_{1+}}
  + O(e^{-\sigma_{1\pm}}).
\end{equation}
This shows that the limit torus approaches to a geodesic on the
initial surface $(\Sigma,q_{ab})$ as the string thickness
approaches to zero, which is the similar situation to
Eq.(\ref{Frolov-Israel-Unrhu-kappa}).
Eq.(\ref{kappa-inside-thin-string}) also shows that $\kappa=1/a$
when the outside deficit angle vanishes ($\eta=1$). 
This is the ``acceleration'' of the loop with the radius $a$ on a
flat space, i.e., a test string. 
Thus, only by introducing the string thickness, we have obtained
the natural behavior of $\kappa$ which includes the case of a
conical singularity and that of a test string. 
Hence, we may say that the zero thickness of a string is the
essential reason of the geodesic properties obtained in
Ref.\cite{Unruh-Hayward-Israel-McManus}.

%*************************************************************

Though the matter distribution in Sec.(\ref{sec:thick-initial})
is too artificial for a gauge string loop, we may say that a
cosmic string loop is not a geodesic if the loop has its finite
proper length and if its energy density does not diverge at the
core of the string ($\sigma=\infty$).
From the metric (\ref{physical-three-metric}), the proper length
of the Killing orbit of $\varphi^{a}$ is $2\pi a\phi^{2}\sinh\sigma$. 
The core $\sigma=\infty$ has a finite proper length only when
$\phi=O(e^{-\sigma/2})$. 
In this case, the right hand side of
Eq.(\ref{hamiltonian-constraint-2}) behaves
\begin{equation}
  \label{order-of-Tpp}
  - 2\pi G \phi^{5} T_{\perp\perp} =
    O(e^{-\frac{5}{2}\sigma}) \times T_{\perp\perp}. 
\end{equation}
Together with the Hamiltonian constraint
(\ref{hamiltonian-constraint-2}), Eq.(\ref{order-of-Tpp}) yields
that the energy density $T_{\perp\perp}$ of matter fields does
not contribute to $O(e^{-\frac{1}{2}\sigma})$ nor
$O(e^{-\frac{3}{2}\sigma})$ of the conformal factor $\phi$
unless $T_{\perp\perp}$ diverges at $\sigma=\infty$. 
When we solve Eq.(\ref{hamiltonian-constraint-2}), we should
impose the boundary conditions (i), (ii) and $\eta=1$ from the
regularity at the limit torus. 
Then, the asymptotic form of $\phi_{+}$ in
Eq.(\ref{phi+phi-asympt}) is true up to the order of
$O(e^{-\frac{5}{2}\sigma})$ even if we consider the smooth
distribution of $T_{\perp\perp}$.
Further, $a_{0}$ and $a_{1}$ in Eq.(\ref{phi+phi-asympt}) remain
constants. 
These constants $a_{0}$ and $a_{1}$ are determined by the
boundary conditions at the surface of the matter distribution or
the asymptotically flatness at the infinity as in our torus shell
model. 
Then, we obtain Eq.(\ref{kappa-vacuum}), again. 
Thus, the limit torus $\sigma=\infty$ must not be a geodesic
even if the distribution of $T_{\perp\perp}$ is smooth and a
loop string has its finite proper length.

%*************************************************************

%%%%%%%%%%%%%%%%%%%%%%%%%%%%%%%%%%%%%%%%%%%%%%%%%%%%%%%%%%%%%%%%%%%%%%%
\section{Loop cosmic string and linear perturbation}
\label{sec:liniear}
%%%%%%%%%%%%%%%%%%%%%%%%%%%%%%%%%%%%%%%%%%%%%%%%%%%%%%%%%%%%%%%%%%%%%%%

%*************************************************************

In the last section, we have seen a thick loop energy
distribution is not a geodesic on the initial surface, while a
loop conical singularity is. 
The example in the last section also shows that the initial data
derived in Ref.\cite{Frolov-Israel-Unrhu} may be regarded as the
gravitational field produced a smooth matter distribution. 
Further, as mentioned in the Introduction
(Sec.\ref{sec:Introduction}), this also suggests that, for the
fixed loop radius, there
exists a criterion in the string thickness of the 
validity of a linear gravity. 
We call ``{\it critical thickness}''. 
This critical thickness can be evaluated by the discrepancy of a
linear gravity from the initial data in the previous section.

%*************************************************************

The solution to the linearized Hamiltonian constraint is derived
by choosing the line element on $\Sigma$ so that  
\begin{equation}
  \label{Newton-line-element}
  ds^{2} = \left(1 - \frac{1}{2}\phi_{N}\right)^{4} ds^{2}_{E}
  \sim (1 - 2 \phi_{N})ds^{2}_{E} .
\end{equation}
The linearized Hamiltonian constraint
(\ref{hamiltonian-constraint-2}) is given by 
\begin{equation}
  \label{Newton-eq}
  \Delta\phi_{N} = 4\pi G T_{\perp\perp} 
\end{equation}
in the leading order of $\phi_{N}$, where $\Delta$ is the
Laplacian on the flat space. 
This is just Newton's equation of gravity.
However, we should comment that cosmic string does not generate
the usual Newton potential in the linearized Einstein gravity
because of its huge tension. 
We must note that $\phi_{N}$ in (\ref{Newton-eq}) does not have
the usual meaning of the Newton potential.

%*************************************************************

We consider the matter energy density distribution given by
$T_{\perp\perp}=\tilde{\mu}\delta(\rho-a)\delta(z)$.
$\tilde{\mu}$ is the line energy density of the string and
$\delta(\rho-a)\delta(z)$ is the $\delta$-function whose support
is on the loop with radius $a$ in the equatorial plane
$z=0$. 
The solution to Eq.(\ref{Newton-eq}) is given by  
\begin{equation}
  \label{Newton-pot}
  \phi_{N} = - \frac{\alpha a}{\sqrt{(\rho+a)^{2}+z^{2}}}K(k),
  \quad
  k^{2} = \frac{4\rho a}{(\rho+a)^{2}+z^{2}},
\end{equation}
where $K(k)$ is the first class complete elliptic integral. 
To derive this solution (\ref{Newton-pot}), we identified the
line energy density $\tilde{\mu}$ with the Hiscock mass
$\alpha/4G$\cite{Frolov-Israel-Unrhu}.

%*************************************************************

\begin{figure}[htbp]
  \begin{center}
    \leavevmode
    \epsfxsize=0.85\textwidth
    \epsfbox[0 0 797 368]{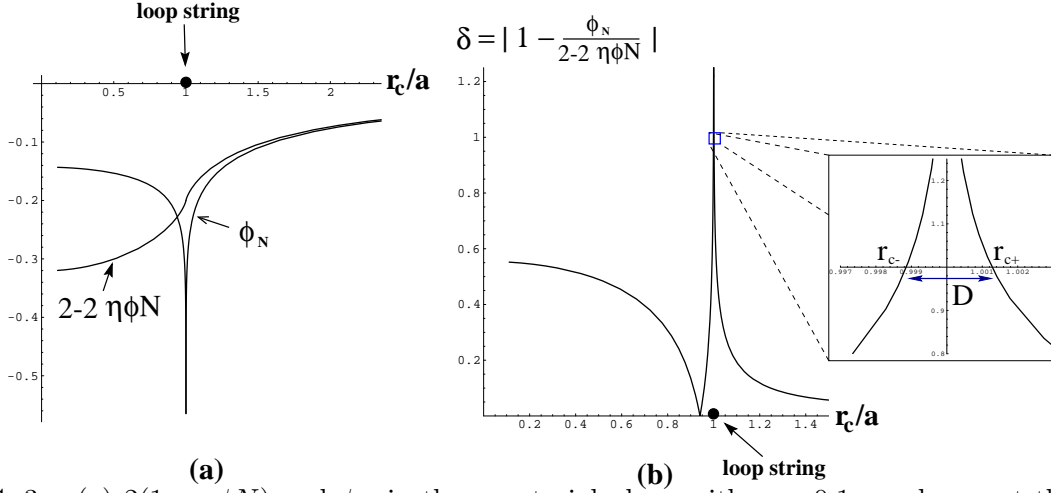}
    \caption{
    (a) $2(1-\eta\phi N)$ and $\phi_{N}$ in the equatorial plane
    with $\alpha=0.1$ are drawn at the same circumference radius
    $r_{c}$ of the Killing orbit of $\varphi^{a}$ normalized by
    the loop radius $a$.
    (b) From $2(1-\eta\phi N)$ and $\phi_{N}$, we obtain the
    profile of $\delta = |1-\phi_{N}/(2-2\eta\phi N)|$. 
    }
    \label{fig:fig3}
  \end{center}
\end{figure}

Comparing with Eq.(\ref{asymptotic-metric}) and
Eq.(\ref{Newton-line-element}), we easily see that $2(1-\eta\phi 
N)$ coincides with $\phi_{N}$ in the asymptotically flat region
$r\rightarrow\infty$. 
As an example, $2(1-\eta\phi N)$ and $\phi_{N}$ on the
equatorial plane with $\alpha=0.1$ are shown in
Fig.\ref{fig:fig3}. 
We evaluate the validity of the linear perturbation theory with
the flat space background by
\begin{equation}
  \label{delta-def}
  \delta := \left|1-\frac{\phi_{N}}{2(1-\eta\phi N)}\right|
\end{equation}
at the same circumference radius of the Killing orbit of
$\varphi^{a}$. We evaluate $\delta$ on the equatorial plane
($z=0$ or $\psi=0,\pi$).
The discrepancy $\delta$ diverges at the locus of the loop due
to the divergence in $\phi_{N}$:
\begin{equation}
  \label{log-diverge-in-PN}
  \phi_{N} \sim \frac{\alpha}{2} \ln
  \frac{\sqrt{(\rho-a)^{2}+z^{2}}}{2a},
\end{equation}
while $\delta\rightarrow 0$ in the asymptotically flat
region. 
This means that $\delta$ becomes order of unity at some points. 
These points give the ``critical thickness''.
If the string thickness is smaller than this critical thickness,
we may regard that the linear perturbation theory on a flat
space fails to describe gravity of a loop string.

%*************************************************************

\begin{figure}[htbp]
  \begin{center}
    \leavevmode
    \epsfxsize=0.55\textwidth
    \epsfbox[17 128 516 620]{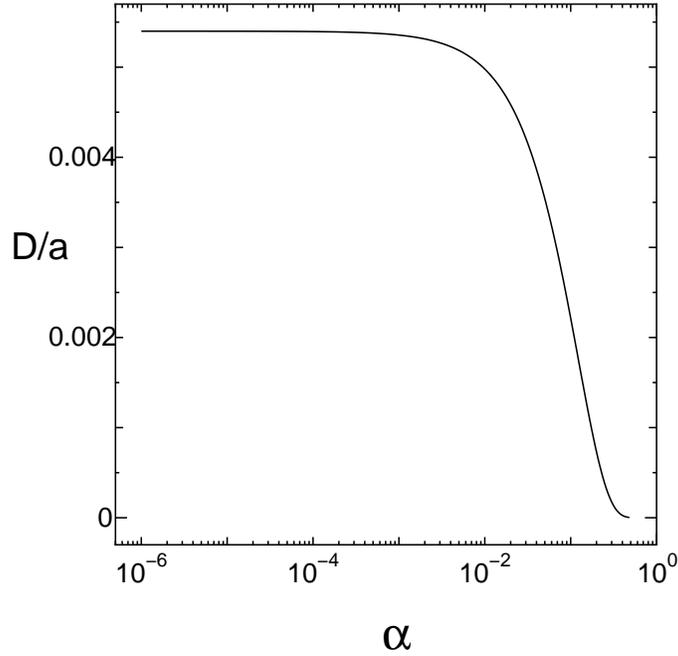}
    \caption{
    The criterion $D$ of a cosmic string thickness for each
    deficit angle $\alpha$ is shown. 
    $a$ is the loop radius. 
    We may say that the linear perturbation theory around flat
    Euclid space fails to describe the gravity of a string when
    the string thickness is smaller than $D$.
    }  
    \label{fig:fig4}
  \end{center}
\end{figure}

Now, we evaluate this criterion. 
There are two circumference radii on the equatorial plane at
which $\delta=1$. 
We denote these by $r_{c\pm}$. 
$r_{c-}$ is in the region $\rho<a$ ($\psi=\pi$) and $r_{c-}$ is
in $\rho>a$ ($\psi=0$). 
We evaluate the critical thickness in the string thickness by
$D:=r_{c+}-r_{c-}$. 
If a loop string has the larger thickness than $D$, we may
regard that the linear perturbation with a flat space background
is plausible.  
The critical thickness $D$ for each deficit angle are shown in
Fig.\ref{fig:fig4}.

%*************************************************************

Fig.\ref{fig:fig4} shows that the linear perturbation theory
with a flat space background is plausible when its thickness is
larger than $\sim 5\times10^{-3} a$ for $\alpha=10^{-6}$ 
(GUT string case).  
This critical thickness is also roughly estimated as follows:
The divergence in $\delta$ is due to the logarithmic one in
$\phi_{N} \sim (\eta-1)/(2\eta) \ln (D/4a)$. 
Since $\phi_{N}<2(1-\eta\phi N)<0$ and $2(1-\eta\phi N) \sim
2(1-\eta)$ near the loop, we obtain $D \sim 4 a e^{-8\eta} \sim
1.4 \times 10^{-3} a$ for $\alpha\sim 10^{-6}$. 
Though this estimation is about one quarter of the result shown
in Fig.\ref{fig:fig4}, it will be due to higher order
corrections in $\phi_{N}$ or $2(1-\eta\phi N)$.

%%%%%%%%%%%%%%%%%%%%%%%%%%%%%%%%%%%%%%%%%%%%%%%%%%%%%%%%%%%%%%%%%%%%%%%
\section{Summary and Discussion}
\label{sec:disucssion}
%%%%%%%%%%%%%%%%%%%%%%%%%%%%%%%%%%%%%%%%%%%%%%%%%%%%%%%%%%%%%%%%%%%%%%%

%*************************************************************
%summary...
%

We have considered the initial data of gravity produced by a
loop string and showed that the core of a thick string is 
not a geodesic on the initial surface, while a loop conical
singularity is. 
To see this, we first consider an example in which the matter
concentrates on the torus surface. 
This geodesic property is unchanged if a loop string has its finite
proper length and the matter energy density does not diverge. 
Then, we may say that the geodesic property obtained in
Ref.\cite{Unruh-Hayward-Israel-McManus} should not be taken
seriously when we consider the dynamics of cosmic strings in the 
early universe, because realistic cosmic strings have their own
finite thickness.

%*************************************************************

Further, using the above example, we have seen that the loop
approaches to a geodesic on the initial surface in the zero
thickness limit.
This behavior in our example also suggests that the metric
perturbation on a flat space background loses its validity when
a string is sufficiently thin.  
We also considered the critical thickness which is the criterion
of the validity of the linear gravity around a flat space
background. 
We evaluate this criterion using the circumference radii at
which the discrepancy $\delta$ defined by Eq.(\ref{delta-def})
becomes unity. 
Fig.\ref{fig:fig4} shows that a linear perturbation is plausible
when the string thickness is larger than $5\times10^{-3}a$ where
$a$ is the string bending scale.

%*************************************************************

It is not clear from the arguments in this paper whether all
behaviors of the ``acceleration'' $\kappa$ of the Killing orbit
in our example are maintained in generic situations.
To confirm this, we must solve the Hamiltonian constraint
(\ref{hamiltonian-constraint-2}) with smooth matter energy
distribution, numerically. 
However, we expect that these are also true even when the matter
distribution is smooth because the coefficients $a_{0}$ and
$a_{1}$ in Eq.(\ref{kappa-vacuum}) are completely determined by
the regularity at the limit torus and the asymptotically conical
structure.

%*************************************************************

Fig.\ref{fig:fig3}(a) also shows that there is the discrepancy
of $\phi_{N}$ and $2(1-\eta\phi N)$ in the region near loop axis
$z=0$. 
This will be due to the degree of freedom of gravitational waves
on the initial surface. 
This speculation is also suggested the fact that the usual
Newton potential does not generated by a cosmic string because
of its huge tension as commented in the previous section.
To confirm this speculation, we must solve the time evolution
from this initial data. 
However, $\delta$ in this region is smaller than unity and this
discrepancy will be described by the linear perturbation with
the Minkowski spacetime background. 
So this discrepancy is not important for the above criterion.

%*************************************************************

In Ref.\cite{kouchan}, we discussed the dynamics of a
fluctuating infinite string and showed that the displacement of
a string vanishes for a zero thickness string.
These results in Ref.\cite{kouchan} is due to the same situation 
discussed here. 
From the viewpoint of the idealization problem of concentrated
line sources in general relativity, we may say that the
thickness should be remained finite when we investigate the
dynamics of self-gravitating strings and their gravitational
wave emission.

%*************************************************************

We must note that our conclusion here does not contradict to our
conclusion in Ref.\cite{kouchan}: there is no dynamical degree
of freedom of the free oscillations of a Nambu-Goto string
within the first order with respect to its oscillation
amplitude. 
In Ref.\cite{kouchan}, we have carefully excluded the gauge
freedom of perturbations. 
The result obtained here suggests that we may use the linear
perturbation theory with Minkowski background if the string
thickness is much larger than the above critical thickness,
though we must bear in our mind that the effect of Higgs and
gauge fields constructing cosmic string will dominate in a
realistic cosmic string.  
In this case, we have to exclude gauge freedom of the
perturbations carefully as in Ref.\cite{kouchan}. 
This will be our future work.

%*************************************************************

Finally, we must also comment that our arguments here and in
Ref.\cite{kouchan} are those for a string in the situation where
its bending scale is much larger than its thickness. 
So, from the conclusion here, we cannot say anything about the
geometry of near cusps or kinks, which may arise in the complex
dynamics of cosmic strings\cite{Damour-Vilenkin}. 
When cusps or kinks appear by the dynamics from large loop, we
will have to take into account the dynamics of Higgs and gauge
fields. 
This will be not the problem in general relativity but the field
theory itself.

%%%%%%%%%%%%%%%%%%%%%%%%%%%%%%%%%%%%%%%%%%%%%%%%%%%%%%%%%%%%%%%%%%%%%%%
\section*{acknowledgements}

The author thanks Prof. Minoru Omote, Prof. Takashi Mishima and
Prof. Akio Hosoya for their continuous encouragement.

%%%%%%%%%%%%%%%%%%%%%%%%%%%%%%%%%%%%%%%%%%%%%%%%%%%%%%%%%%%%%%%%%%%%%%%
\appendix
%%%%%%%%%%%%%%%%%%%%%%%%%%%%%%%%%%%%%%%%%%%%%%%%%%%%%%%%%%%%%%%%%%%%%%%
\section{Conformally non-flat initial data}
\label{sec:c-non-flat}
%%%%%%%%%%%%%%%%%%%%%%%%%%%%%%%%%%%%%%%%%%%%%%%%%%%%%%%%%%%%%%%%%%%%%%%

%*************************************************************

In this appendix, we consider the conformally non-flat version
of the initial data discussed by Frolov et
al\cite{Frolov-Israel-Unrhu} and show that, even in this
initial data, a thick loop energy distribution is not a geodesic
on the initial surface $\Sigma$ nevertheless a conical
singurality is.

%*************************************************************

The conformally non-flat version of the initial data discussed
by Frolov et al\cite{Frolov-Israel-Unrhu} is characterized by
the conformally transformed metric $\tq_{ab}$
\begin{equation}
  \label{toroidal-nonflat-2}
  \tq_{ab} = a^{2} \left( e^{2\nu} (d\sigma)_{a}(d\sigma)_{b} 
    + e^{2\nu} (d\psi)_{a}(d\psi)_{b} 
    + \sinh^{2}\sigma (d\varphi)_{a}(d\varphi)_{b}\right).
\end{equation}
In this initial data, there are two unknown functions
$\nu(\sigma,\psi)$ and the conformal factor $\phi(\sigma,\psi)$
and the Hamiltonian constraint (\ref{hamiltonian-constraint-2})
is given by
\begin{equation}
  \label{hamil-con-toroical-non-flat}
  \frac{1}{\sinh\sigma} \partial_{\sigma} (\sinh\sigma\partial_{\sigma}\phi)
  + \partial_{\psi}^{2}\phi + \frac{1}{4} \left( 1 +
  (\partial_{\sigma}^{2} + \partial_{\psi}^{2})\nu \right) \phi 
  = - 2 \pi G a^{2}e^{2\nu} \phi^{5} T_{\perp\perp}.
\end{equation}

%*************************************************************

As a vacuum solution to Eq.(\ref{hamil-con-toroical-non-flat})
with a loop conical singularity, 
\begin{eqnarray}
  \label{51}
  \nu &=& \left\{
    \begin{array}{ccc}
      \alpha (\sigma-\sigma_{0}), &\quad & (\sigma>\sigma_{0}),\\
      0, &\quad & (\sigma<\sigma_{0}),
    \end{array}\right. \\
  \label{phi-sol-nonflat-vacuum}
  \phi &=& N^{-1} + \frac{\sqrt{2}}{\pi}
  \sum_{n=0}^{\infty} \frac{\alpha\epsilon_{n}b_{n}}{4-\alpha b_{n}} 
  \frac{P_{n-1/2}(\cosh\sigma_{<})}{P_{n-1/2}(\cosh\sigma_{0})} 
  Q_{n-1/2}(\cosh\sigma_{>}) \cos n\psi
\end{eqnarray}
is derived in Ref.\cite{Frolov-Israel-Unrhu}, where $\sigma_{<}$
and $\sigma_{>}$ denote the lesser and greater of
$(\sigma,\sigma_{0})$ and $b_{n}$ are given by   
\begin{equation}
  \label{67}
  b_{n}=\sinh\sigma_{0}Q_{n-1/2}(\cosh\sigma_{0})P_{n-1/2}(\cosh\sigma_{0}).
\end{equation}
Eq.(\ref{51}) guarantees that the limit torus $\sigma=\infty$ is
a conical singularity and the axis $\sigma=0$ is
not\cite{Frolov-Israel-Unrhu}.

%*************************************************************

In the solution given by Eqs.(\ref{51})-(\ref{67}), we easily
see that the conical singularity at $\sigma=\infty$ is a
geodesic on the initial surface $\Sigma$.  
To see this, we consider the region $\sigma>\sigma_{0}$. 
In this region, Eq.(\ref{51}) yields $\nu = \alpha
(\sigma-\sigma_{0})$ and Eq.(\ref{phi-sol-nonflat-vacuum}) is
given by
\begin{eqnarray}
  \label{phi-near-string}
  \phi &=& \frac{\sqrt{2}}{\pi}
  \sum_{n=0}^{\infty} \frac{4 \epsilon_{n}}{4-\alpha b_{n}} 
  Q_{n-1/2}(\cosh\sigma) \cos n\psi \\
  &=& \frac{4\sqrt{2}e^{-\frac{\sigma}{2}}}{4 - \alpha b_{0}} 
  \left( 1 + \frac{4-\alpha b_{0}}{4-\alpha b_{1}} e^{-\sigma}\cos\psi 
    + O(e^{-2\sigma}) \right) \nonumber.
\end{eqnarray}
The ``acceleration'' of the Killing orbit of $\varphi^{a}$ behaves
\begin{equation}
  \kappa \sim \frac{1}{a} e^{\alpha\sigma_{0}} \frac{(4-\alpha
  b_{0})^{3}}{4-\alpha b_{1}} e^{-\alpha\sigma}
  \propto \left(\frac{d}{a}\right)^{\frac{\alpha}{1-\alpha}},
\end{equation}
where the proper distance $d \sim a e^{-(1-\alpha)\sigma}$ from
the Killing orbit with a finite $\sigma$ to the conical
singularity ($\sigma=\infty$). 
This shows that the loop conical singularity is a geodesic on
$\Sigma$.

%*************************************************************

Here, we show that the limit torus $\sigma=\infty$ is not a
geodesic when we replace the above conical singularity by the
matter distribution with the finite energy. 
For simplicity, we consider a $\delta$-function distribution
of a matter energy density at the torus
$\sigma=\mbox{constant}$, denoted by ${\cal S}$, as in the main
text. 
The metric $q_{ab}=\phi^{4}\tq_{ab}$ on $\Sigma_{-}$ (outside
the torus) is given by Eqs.(\ref{toroidal-nonflat-2}),
(\ref{51}), and (\ref{phi-sol-nonflat-vacuum}). 
On $\Sigma_{+}$, we assume that the metric $q_{ab}$ is given by
$\nu=0$ and Eq.(\ref{vacuum-solution-inside}).

%*************************************************************

The normal vectors $n_{a\pm}$ of $\partial\Sigma_{\pm}$ are
given by
\begin{equation}
  n_{a\pm} = a \phi^{2}_{\pm}(\psi) e^{\nu_{\pm}} (d\sigma_{\pm})_{a}, 
\end{equation}
where $\phi_{\pm}$ and $\nu_{\pm}$ are metric functions $\phi$
and $\nu$ on the boundaries $\sigma=\sigma_{1\pm}$, i.e.,
$\partial\Sigma_{\pm}$, respectively. 
The induced metrics $h_{ab\pm}$ on $\partial\Sigma_{\pm}$ are
given by
\begin{equation}
  h_{ab\pm} = a^{2} \phi^{4}_{\pm} 
  \left( e^{2\nu_{\pm}} (d\psi)_{a}(d\psi)_{b} 
    + \sinh^{2}\sigma_{1\pm} (d\varphi)_{a}(d\varphi)_{b}\right).
\end{equation}
The trace of the extrinsic curvatures $\kappa^{\;\;c}_{c\pm}$ of
$\partial\Sigma_{\pm}$ in $\Sigma_{\pm}$ are given by
\begin{equation}
  \label{extrinsic-curvature-formula}
  \kappa_{c\pm}^{\;\;c} = \left.
  - \frac{\partial_{\sigma}(\phi^{2}e^{\nu})}{a\phi^{4}e^{2\nu}}
  - \frac{1}{a\phi^{4}e^{\nu}\sinh\sigma} 
  \partial_{\sigma}(\phi^{2}\sinh\sigma). \right|_{\sigma=\sigma_{1\pm}}.
\end{equation}

%*************************************************************

We only consider the case $\sigma_{1-}>\sigma_{0}$. 
Then we may use $\nu_{+}=0$, $\nu_{-}=\alpha(\sigma_{1-}-\sigma_{0})$. 
$\phi_{-}$ is given by Eq.(\ref{phi-near-string}) with
$\sigma=\sigma_{1-}$ and $\phi_{+}$ is given by
Eq.(\ref{vacuum-solution-inside}) with $\sigma=\sigma_{1+}$. 
The junction condition (\ref{constraint-junction-intrinsic}) for
the intrinsic metric yields     
\begin{equation}
  \label{intrinsic-junctions-nonflat}
  e^{-\nu_{-}}\sinh\sigma_{1-} = \sinh\sigma_{1+}, \quad
  \phi_{+} = \phi_{-} e^{\frac{\nu_{-}}{2}}.
\end{equation}
As in the conformally flat version, the first condition gives
the relation of $\sigma_{1\pm}$. 
The second condition determines the coefficients $a_{n}$ in
Eq.(\ref{vacuum-solution-inside}):
\begin{equation}
  \label{an_in_cnf-version}
  a_{n} = \frac{4e^{\frac{\nu_{-}}{2}}}{4-\alpha b_{n}}
  \frac{Q_{n-\frac{1}{2}}(\cosh\sigma_{1-})}{Q_{n-\frac{1}{2}}(\sqrt{1
  + e^{-2\nu_{-}}\sinh^{2}\sigma_{1-}})} 
  = \frac{4e^{\frac{\nu_{-}}{2}}}{4-\alpha b_{n}}
  \frac{Q_{n-\frac{1}{2}}(\sqrt{1 +
  e^{2\nu_{-}}\sinh^{2}\sigma_{1+}})}{Q_{n-\frac{1}{2}}(\cosh\sigma_{1+})}.  
\end{equation}
The junction condition (\ref{constraint-junction-extrinsic}) for
the extrinsic curvature $\kappa_{c\pm}^{\;\;c}$ gives the energy
density on the torus shell as in the conformally flat version in
the main text.

%*************************************************************

In this initial data, we easily see that the limit torus
$\sigma=\infty$ in this model is not a geodesic on
$(\Sigma,q_{ab})$ (on $(\Sigma_{+},q_{ab+})$). 
Since the metric on $\Sigma_{+}$ is also given by
Eq.(\ref{inside-physical-metric}) and
Eq.(\ref{vacuum-solution-inside}), the ``acceleration'' $\kappa$
of the Killing orbit of $\varphi^{a}$ at $\sigma=\infty$ is
given by Eq.(\ref{kappa-vacuum}), again. 
Since the coefficients $a_{n}$ in Eq.(\ref{an_in_cnf-version})
are constants determined by the locus of the torus matter shell,
we conclude that the limit torus $\sigma=\infty$ is not a
geodesic on $(\Sigma,q_{ab})$. 
As in the case in Sec.\ref{sec:is-a-loop-string-is-geodesic},
when the matter torus shell is sufficiently thin
($\sigma_{1\pm}\rightarrow\infty$), $\kappa$ behaves
\begin{equation}
  \label{kappa-conformally-nonflat-asympt}
  \kappa \propto \left(\frac{d_{1+}}{a}\right)^{\frac{\alpha}{1-\alpha}},
\end{equation}
where $d_{1+}=a\int^{\infty}_{\sigma_{1+}}d\sigma\phi^{2}$ is
the proper distance from $\sigma=\sigma_{1+}$ to the limit torus
$\sigma=\infty$.
Eq.(\ref{kappa-conformally-nonflat-asympt}) shows that when the
thickness of the string is sufficiently smaller than its
curvature radius ($d_{1+}/a\rightarrow0$), the loop approaches
to a geodesic on $(\Sigma,q_{ab})$. 
These behaviors are same as that obtained in
Sec.\ref{sec:is-a-loop-string-is-geodesic}.

%%%%%%%%%%%%%%%%%%%%%%%%%%%%%%%%%%%%%%%%%%%%%%%%%%%%%%%%%%%%%%%%%%%%%%%

%\end{multicols}
\end{document}